\documentclass[conference]{IEEEtran}

\usepackage{graphicx} 
\graphicspath{{Fig/}}

\usepackage{graphicx}
\usepackage{amsmath,bbm,amssymb,amsfonts,amstext,amsopn,sansmath}
\usepackage{cite}
\usepackage{balance}
\usepackage{url}
\usepackage{epsfig}
\usepackage{setspace}
\usepackage{stmaryrd}
\usepackage{psfrag}	
\usepackage{multirow}
\usepackage{makecell}
\usepackage{float}
\usepackage[process=auto]{pstool}
\usepackage{etoolbox}
\usepackage{algorithm}
\usepackage{algorithmic}
\allowdisplaybreaks
\usepackage{textcomp}
\usepackage{multicol}
\usepackage{xfrac}
\usepackage{colortbl}

\usepackage{clipboard}
\newclipboard{Paper_V1}
 
\usepackage{xcolor}
\makeatletter
\let\myorg@bibitem\bibitem
\def\bibitem#1#2\par{%
	\@ifundefined{bibitem@#1}{%
		\myorg@bibitem{#1}#2\par
	}{%
		\begingroup
		\color{\csname bibitem@#1\endcsname}%
		\myorg@bibitem{#1}#2\par
		\endgroup
	}%
}
\newcommand{\highlightref}[1]{\expandafter\newcommand\expandafter*\csname bibitem@#1\endcsname{blue}}
\makeatother

\highlightref{zheng2019intelligent}

\newcommand{\Trans}{\mathsf{T}}

\def\bs{\mathbf{s}}
\def\by{\mathbf{y}}
\def\bA{\mathbf{A}}

\def\bx{\mathbf{x}}
\def\bn{\mathbf{n}}
\def\bV{\mathbf{V}}

\def\bone{\boldsymbol{1}}
\def\bzero{\boldsymbol{0}}

\def\bA{\mathbf{A}}

\def\bu{\mathbf{u}}

\def\ba{\mathbf{a}}
\def\bM{\mathbf{M}}
\def\bw{\mathbf{w}}
\def\bX{\mathbf{X}}

\def\byrv{\boldsymbol{\mathnormal{y}}}
\def\bxrv{\boldsymbol{\mathnormal{x}}}

\def\sR{\mathbb{R}}
\def\sRp{\sR_{\geq0}}

\def\sQ{\mathcal{Q}}
\def\sM{\mathcal{M}}
\def\sX{\mathcal{X}}

\def\Ex{\mathbb{E}} 
\def\Vx{\mathbb{V}}

\def\cA{\mathcal{A}}
\def\cAc{\mathcal{A}^{\rm c}}

\def\cM{\mathcal{M}}

\newcommand{\diag}{\mathrm{diag}}

\newcommand{\thr}{\mathrm{thr}}

\newcommand{\argmax}{\mathrm{argmax}}

\newcommand{\rls}{\mathrm{rls}}

\newtoggle{OneColumn}
\toggletrue{OneColumn}

\newtoggle{arXiv}
\togglefalse{arXiv}

\usepackage{lipsum}
\def\@IEEEBIOphotowidth{1cm}    
\def\@IEEEBIOphotodepth{1cm}   
\def\@IEEEBIOhangwidth{1.2cm}    
\def\@IEEEBIOhangdepth{1.2cm}    

\EndPreamble
\begin{document}
\title{Compressive Sensing-Based Recovery of Molecular Mixtures with Cross-Reactive Receptor Arrays \vspace{-0.15cm}}

\author{Vahid Jamali$^{\ast,\circ}$, Helene M. Loos$^{\bullet,\star}$, Andrea Buettner$^{\bullet,\star}$,  Robert Schober$^{\bullet}$, and H. Vincent Poor$^{\ast}$\\
	$^{\ast}$Princeton University,
	$^{\circ}$Technical University of Darmstadt,
	$^{\bullet}$Friedrich-Alexander-Universit\"at Erlangen-N\"urnberg,\\
	$^{\star}$Fraunhofer Institute for Process Engineering and Packaging IVV
	  \vspace{-0.2cm}
}

\maketitle

\begin{abstract}
In this paper, we propose a novel concept for engineered molecular communication (MC) systems inspired by animal olfaction. We focus on a multi-user scenario where  transmitters employ unique mixtures of different types of signaling molecules to  convey their messages to  a central receiver, which is equipped with an array comprising $R$ different types of receptors  to detect the emitted molecule mixtures.  The hardware complexity of an MC system employing \textit{orthogonal} molecule-receptor pairs would linearly scale  with the number of signaling molecule types $Q$ (i.e., $R=Q$). Natural olfaction systems avoid such high complexity by employing  arrays of \textit{cross-reactive} receptors, where each type of molecule activates multiple types of receptors and each type of receptor is predominantly activated by multiple types of molecules albeit with different activation strengths. For instance, the human olfactory system is believed to discriminate several thousands of chemicals using only a few hundred receptor types, i.e., $Q\gg R$. Motivated by this observation, we first develop an end-to-end MC channel model that accounts for the key properties of olfaction. Subsequently, we formulate the molecule mixture recovery as a convex compressive sensing (CS) problem which can be efficiently solved via available numerical solvers. Our simulation results confirm the efficiency of the proposed CS problem for the recovery of the  molecular mixture signal and quantify the system performance for various system parameters. 
\end{abstract}

\section{Introduction}

In electromagnetic- (EM-) based communication systems, exploiting different \textit{frequency resources} is essential for an efficient use of the communication channel. In  molecular communication (MC) systems,   the different \textit{types of signaling molecules} used for communication are analogous to different frequency resources \cite{jamali2019channel}. This has lead researchers to propose molecular shift keying (MSK) modulation, which embeds information in the identity of signaling molecules \cite{kuscu2019transmitter,kuran2011modulation,jamali2018diffusive,chen2020generalized}. Thereby, ideally, by employing a large number of different types of signaling molecules, the MC channel is able to offer high throughput, despite the slow propagation speed of molecules. However, the MSK-based MC designs proposed so far in the literature \cite{kuscu2019transmitter,kuran2011modulation,jamali2018diffusive,chen2020generalized} employ only a small number of different types of molecules, denoted by $Q$,  and are not scalable for large $Q$ due to the following limitations: 
\textbf{First}, it is implicitly assumed in MSK modulation that for each adopted signaling molecule type, the receiver (Rx) is equipped with a corresponding specifically-tuned receptor type. This implies that the Rx architecture complexity linearly increases with $Q$ which limits the feasibility of this Rx architecture for large $Q$. 
\textbf{Second}, the ideal assumption that each receptor type \textit{only} responds to the corresponding signaling molecule type  simplifies the design of the MC system but  may not be valid in practice \cite{tisch2010nanomaterials}. Most receptors respond to different chemical substances  albeit with different strengths. 
For future reference, we refer to  MC systems that employ a Dedicated
Receptor type for EAch Molecule type  as DREAM systems. 

\begin{figure}[t]
	\centering
	\includegraphics[width=0.8\columnwidth]{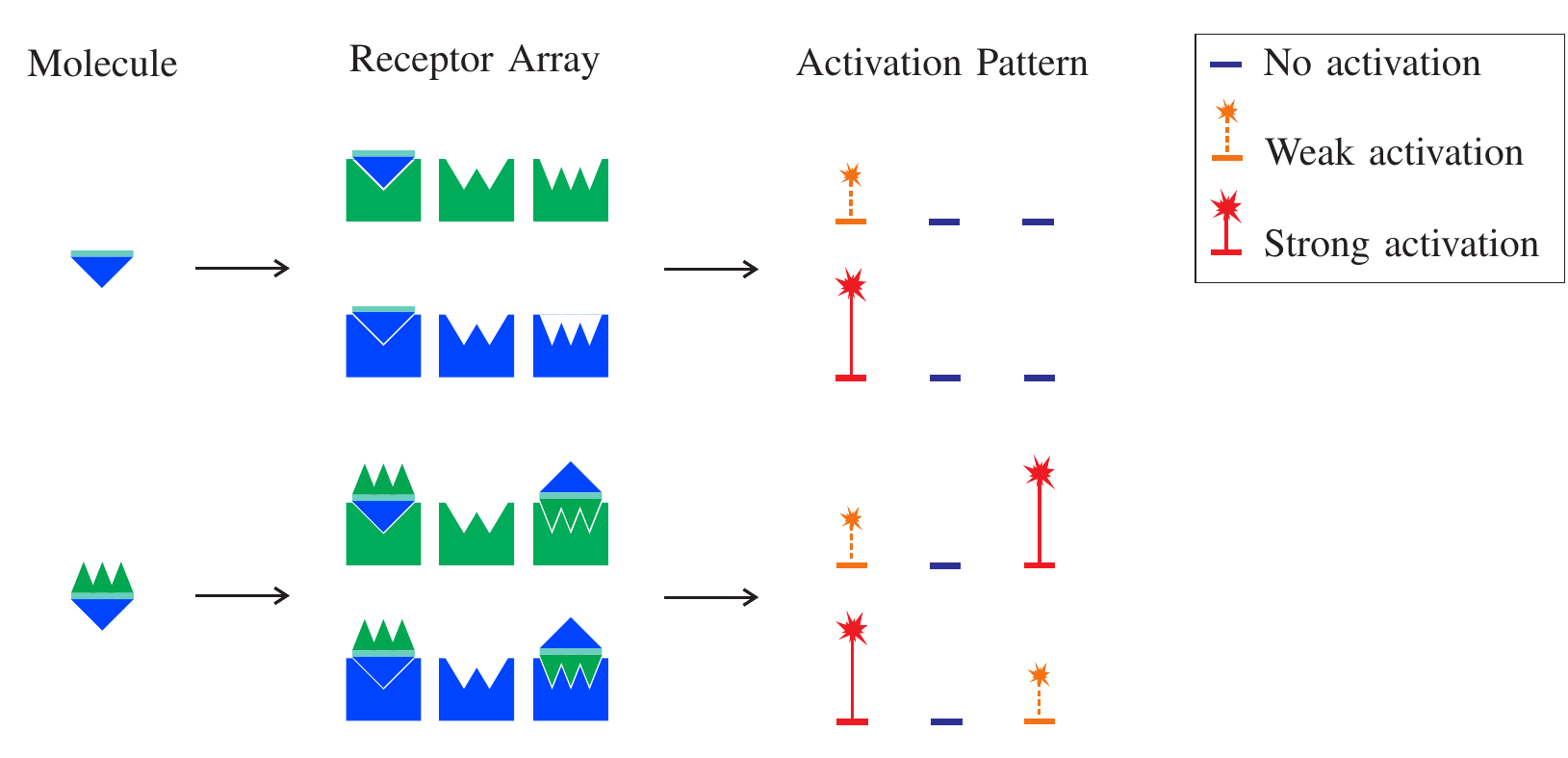}\vspace{-0.3cm}
	\caption{Schematic illustration of the cross-reactive receptor arrays of  olfactory systems known as \textit{shape-pattern theory of olfactory perception} \cite{buck2005unraveling}. In this illustration, each molecule type activates those receptor types that have a matching geometrical shape and the strength of activation is enhanced if they have identical color. } \label{Fig:Reception} \vspace{-0.3cm}
\end{figure}

Fortunately, nature offers several solutions for the concurrent detection of a large number of  different types of molecules, which can be found in the chemosensory systems of different organisms. In fact, for airborne chemicals, the olfactory systems of mammals and insects can detect molecular information in an efficient manner and have vital functions in, e.g., navigating, foraging, and reproduction  \cite{buck2005unraveling,su2009olfactory,kaupp2010olfactory}. For instance, it
is estimated that humans are able to perceive $\sim 10^4-10^5$ different chemicals as having  distinct odors using only $\sim 400$ different types of olfactory receptors (ORs)\footnote{The Nobel Prize in Physiology and Medicine in 2004 was awarded jointly to Richard Axel and Linda B. Buck for their discoveries of ``odorant receptors and the organization of the olfactory system`` \cite{buck2005unraveling}.}  \cite{buck2005unraveling}.  Unlike DREAM systems where each receptor type is assumed to be narrowly tuned to its respective type of signaling molecule, in these natural systems, the majority of the receptors are broadly tuned to multiple types of signaling molecules while most signaling molecule types have affinity with multiple types
of receptors. Therefore, the odor discriminatory capacity of the olfactory system stems from an array of \textit{cross-reactive} receptor types, which extracts the information regarding the presence of  molecules and encodes it into the \textit{activation pattern} of the ORs, see Fig.~\ref{Fig:Reception}.

The main objective of this paper is to investigate the potential benefits of exploiting the properties of natural olfaction for the design of engineered MC systems. In particular, this paper makes the following  contributions: 
\begin{itemize}
		\item Inspired by natural odor mixture communication \cite{thomas2014perception}, we introduce a new modulation scheme where information is embedded in the mixtures of different types of signaling  molecules, referred to as molecule mixture shift keying (MMSK) modulation. 
	\item We develop a communication-theoretic model that relates the molecule mixture signal to the received array signal. This model captures the key properties of olfaction including cross-reactive receptor arrays (CRRAs). 
	\item Finally, we formulate the molecule mixture recovery at the Rx as a compressive sensing (CS) problem which explicitly exploits the knowledge of the MMSK modulation alphabets used by the transmitters (Txs). Due to the specific structure of the MMSK modulated signal and the non-linearity of the received array signal, the formulated problem introduces new challenges, which are tackled in this paper. 
\end{itemize}
To the best of the authors' knowledge, the olfactory-inspired engineered MC system employing a CRRA has been first reported in this paper and its extended version \cite{jamali2022olfaction}. We further note that the multiple-input multiple-output (MIMO) MC systems studied in the literature \cite{meng2012mimo,koo2016molecular} employ multiple receptors of the \textit{same} type and a \textit{single} type of molecule and their generalization to MSK modulation suffers from the same limitations as the DREAM MC systems discussed above. 

\textit{Notation:} Bold small letters (e.g., $\bx$) and bold capital letters  (e.g., $\bX$) denote vectors and matrices, respectively. Sets are shown by calligraphic letters (e.g., $\sX$). $\bX^{\Trans}$ represents the transpose of the matrix $\bX$ and $[\bX]_{n,m}$ denotes the entry of the $n$-th row and $m$-th column of the matrix $\bX$ where sub-index $m$ is dropped for vectors. We use $\by = |\bx|$ to denote the element-wise absolute value (i.e., $[\by]_{n}=|[\bx]_{n}|$). $\|\bx\|_p$ denotes the vector $p$-norm and $\bone_{n}$ and $\bzero_n$ are the all-one and all-zero vectors of size $n$, respectively. $|\sX|$ denotes the cardinality of the set $\sX$.  $\sR$ and $\sRp$ are the sets of real and non-negative real numbers, respectively. ${\rm Pois}(\lambda)$ represents a Poisson random variable (RV) with mean $\lambda$.  $\Ex\{\bX\}$ and $\Vx\{\bX\}$ denote the expectation and variance, respectively, of the entries of the matrix  $\bX$.

\section{Communication-Theoretic End-to-End Channel Model for Molecular Mixture Signaling}\label{sec:channel}

\begin{figure}[t]
	\centering
	\includegraphics[width=0.7\columnwidth]{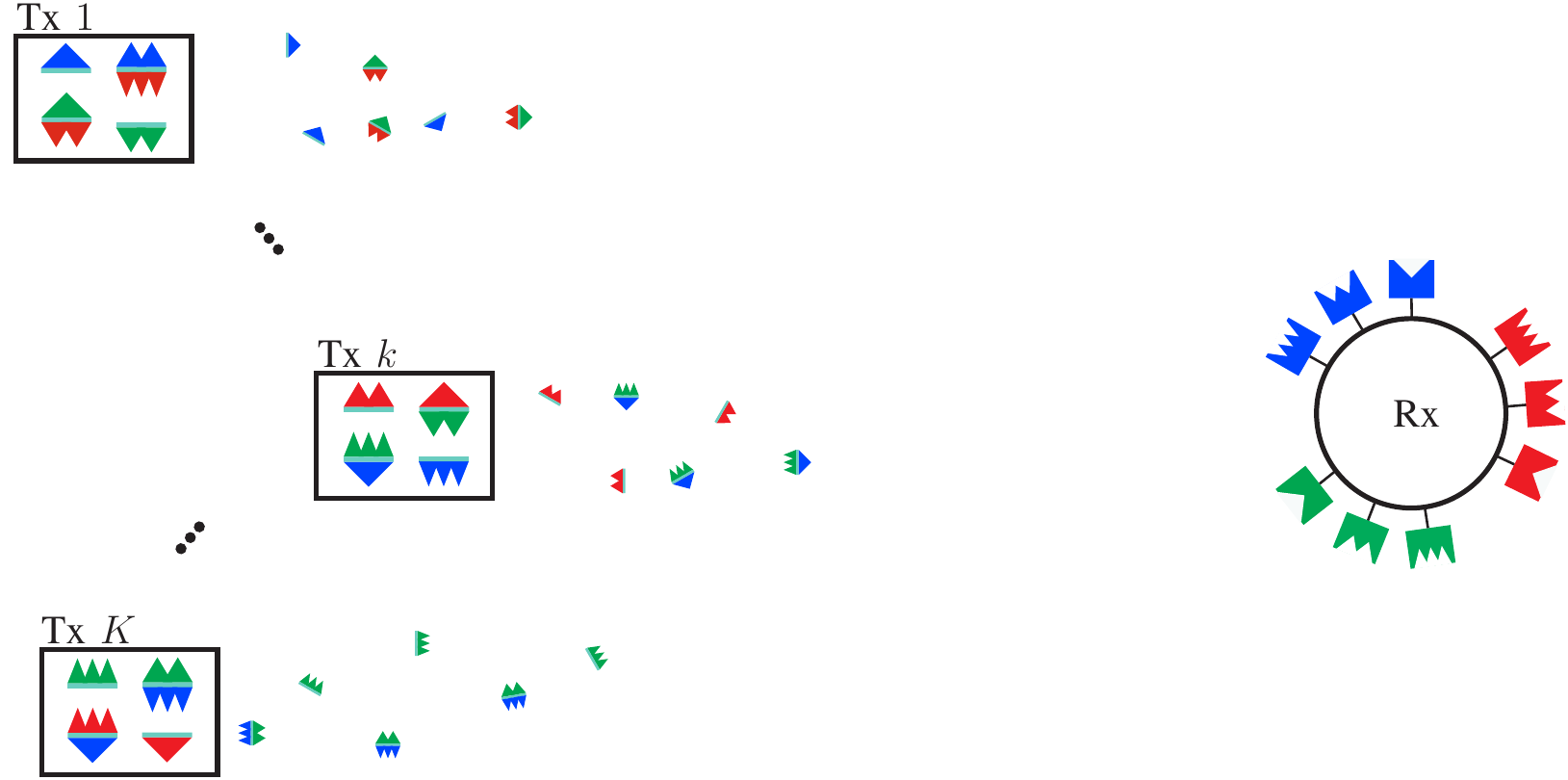}\vspace{-0.1cm}
	\caption{Proposed multi-user MC system consisting of several Txs that employ MMSK modulation to convey their messages to an Rx equipped with a CRRA.   \vspace{-0.3cm}} \label{Fig:MultiuserSystem} \vspace{-0.3cm}
\end{figure}

We consider a multi-user MC system consisting of $K$ Txs  sending  messages to an Rx, e.g., as a model of a sensor network, see Fig.~\ref{Fig:MultiuserSystem}. We assume that the Txs employ MMSK modulation, namely each Tx releases a mixture of different molecule types to represent its message. We assume each Tx sporadically accesses the MC channel in a random access manner since it has only occasionally a small amount of data to transmit (e.g., reporting a change in temperature measured by a  sensor) which does not justify the establishment of synchronous communication. The released molecules propagate through the MC channel (e.g., via diffusion, turbulent flow, etc.) and some reach the Rx. The Rx is equipped with an array of $R$ types of cross-reactive receptors and recovers the transmitted messages and the identity of the corresponding Txs by processing the array signals. The objective of this section is to develop an end-to-end channel model that relates the  mixture signals released by the Txs to the  received array signals at the Rx. To do so, in the following, we first review the basic properties of olfaction, which form the basis for the proposed communication-theoretic channel model.

\subsection{Key properties of olfaction}\label{sec:olfprop} 
Some of the key properties of natural olfactory systems are discussed in the following: 
\begin{itemize}
	\item \textbf{P1 (cross-reactive receptor array):} The main reason for the high discriminatory capacity of olfactory systems is their CRRA structure. In other  words, one type of molecule  may activate multiple types of receptors and conversely, one type of receptor can be activated by different types of molecules \cite{buck2005unraveling,kaupp2010olfactory,mori1995molecular}.
	\item \textbf{P2 (sparsity):} Although the olfactory system is
	able to distinguish thousands of different molecule types, the number of molecule types that it can simultaneously identify is limited. For instance, various experiments demonstrated that humans can hardly distinguish $3-4$ odorants in a mixture that contains up to $8$ odorants \cite{thomas2014perception}. 
	\item \textbf{P3 (concentration):} Higher concentrations of molecules activate more types of receptors because the activation of some types of receptors  is concentration dependent requiring a minimum concentration of molecules to proceed with detection \cite{su2009olfactory,bolding2017complementary}. 
	\item \textbf{P4 (inhibition):} A given type of molecules can activate some types of receptors
	and inhibit others, whereas an individual receptor can be activated by some types of molecules and inhibited by others \cite{su2009olfactory}. While various examples of receptor inhibition have been discussed in the literature, see, e.g., \cite{hallem2004molecular,pfister2020odorant},  inhibition  is less common (or at least not sufficiently understood) compared to  excitation/activation  particularly in vertebrates \cite{kaupp2010olfactory}.  
	\item \textbf{P5 (noise):} OR neurons are noisy and spontaneously active (i.e., they fire action potentials even in the absence of an odorant). For instance, the OR neuron of a Drosophila fly fires $\sim8$ spikes/s in the absence of odor \cite{wilson2013early}. The aggregation of the output signals of all OR neurons of the same type yields thousands of noisy baseline spikes/s. 
	\item \textbf{P6 (sensitivity enhancement):} All OR neurons of the same type send their signals to only few ($\sim2-4$) spherical structures, called glomeruli, in the olfactory bulb. This signal aggregation enhances the detection sensitivity and the signal-to-noise ratio (SNR) of the subsequent signal sent to the olfactory cortex  by averaging out the uncorrelated noises across the distributed receptors~\cite{laurent1999systems}.	
\end{itemize}


\subsection{End-to-end channel model} 
The proposed end-to-end channel model accounts for the release, propagation, and reception mechanisms of the system. In particular, as  channel inputs, we consider  mixtures of different types of signaling molecules released into the channel by the Txs, where each molecule mixture coveys a certain message (e.g., a change in the temperature measured by a sensor). Let $\sQ$ and $\sM$ denote the sets of all molecule types and molecule mixtures used by the Txs, respectively, where $Q\triangleq|\sQ|$ and $M\triangleq|\sM|$. The transmit signal vector is defined as $\bs=[s_1,\dots,s_{M}]^{\Trans}\in\{0,1\}^{M}$ where  $s_m=1$ if  molecule mixture $m$ is released, otherwise $s_m=0$. We assume that the release times of the Txs are not synchronous or coordinated. We further assume that the molecules propagate independently from each other and the signaling molecules do not react with each other \cite{jamali2019channel}. As channel output,  we consider the aggregated signal of the receptors of the same type, e.g., the output signals of the glomeruli. Let $\by=[y_1,\dots,y_R]^{\Trans}\in\sRp^R$ denote the received signal vector, where $y_r$ is the aggregated output signal of the type $r$ receptors. 
We focus on a short time window of the received signal during which each mixture may be released only once\footnote{We focus on a single snapshot of the array  signal due to the assumption of  asynchronous and sporadic multiple access, which is motivated by  natural olfactory systems, where a large number of  odors has to be recognized for animal survival, however, in practice, only a few odors may be simultaneously present  \cite{thomas2014perception}. The more challenging scenario, where Txs continuously access the MC channel,  introduces inter-symbol and multi-user interference and may lead to a non-sparse received signal, which severely limits the number of detectable  mixtures. Therefore, the study of this scenario is beyond the scope of this paper but  constitutes an interesting direction for future research.}. This leads to the following communication model for relating channel input $\bs$ and output $\by$:
\begin{IEEEeqnarray}{lll}
	\text{Release:}\quad &\bu=  N_{\rls} \bM\bs\label{Eq:Release} \\ 
	\text{Propagation:}\,\, &\bx = {\rm Pois}\big(\bV\bu\big) \label{Eq:Propagation} \\	
	\text{Reception:}\quad	&\by = f\big(\bA \bx +\bn\big). \label{Eq:Reception}
\end{IEEEeqnarray}
\textbf{Release model:} In \eqref{Eq:Release}, $\bu=[u_1,\dots,u_Q]^\Trans\in\sRp^{Q}$, where $u_q$ is the number of released  type $q$ molecules and $N_{\rls}$ is the total number of  molecules that each released mixture contains. 
Moreover, $\bM\in[0,1]^{|\sQ|\times|\sM|}$ is called the molecule-mixture construction matrix, where $[\bM]_{q,m}$ determines the fraction of molecule mixture $m$ that is composed of type $q$ molecules. The sparsity property \textbf{P2} can be enforced by assuming each molecule mixture contains only few types of molecules, i.e., each column of $\bM$ contains few non-zero entries. 

\noindent
\textbf{Propagation model:} In \eqref{Eq:Propagation},  $\bx \in\sRp^{Q}$ denotes the vector of the \textit{total number} of molecules reaching the sensing volume of the Rx (e.g., the mucus layer in the nose) at the sampling time.  We assume that the molecules within the sensing volume either activate a receptor or are degraded so that they cannot activate receptors in the subsequent sampling intervals. This leads to a model that is conceptually similar to the absorbing Rx model widely adopted in the literature, for which the number of received molecules is known to follow a Poisson distribution when $N_{\rls}$ is large \cite{jamali2018diffusive}. Moreover, in \eqref{Eq:Propagation}, $\bV = \diag(v_1,\dots,v_Q)\in\sRp^{Q}$ is a diagonal matrix with diagonal entry $v_q$ denoting the channel response for type $q$ molecules, namely the fraction of molecules reaching the Rx at the sampling time. The specific shape of $v_q(t)$ depends on the propagation mechanism (e.g., diffusion and laminar or turbulent flow), the Tx-Rx distance, and the properties of the signaling molecules (e.g., their size) \cite{jamali2019channel}. 

\noindent
\textbf{Olfactory-inspired reception model:} In \eqref{Eq:Reception},  $\bA\in\sR^{R\times Q}$ denotes the receptor-molecule affinity matrix, where $[\bA]_{r,q}$ determines the strength of the aggregated signal generated by all type $r$ receptors in the presence of a unit concentration of  type $q$ molecules. The cross-reactive property \textbf{P1} is manifested in multiple non-zero elements in each row and column of $\bA$. Examples  of measurement results for the affinity matrix can be found in \cite[Fig.~6]{malnic1999combinatorial} and \cite[Fig.~4]{sicard1984receptor} for a given set of odor molecules and ORs.  Moreover, $f(\cdot)$ is the receptor activation function that is applied to each element of the input vector and is used to model the concentration-dependent non-linearity of the array, cf. property \textbf{P3}. For example, if $f(\cdot)$ is chosen as a rectified linear unit (ReLU) activation function or a SmoothReLU (softplus) activation function, it can model property \textbf{P3} by generating a signal only if the concentration is above a certain threshold. In this paper, we focus on a simple ReLU activation function\footnote{Other activation functions have been considered in the literature, too. For instance,   \cite{qin2019optimal} uses the  nonlinear function $f(x)=\sfrac{x}{(1+x)}$  to model the limited dynamic range of sensors; \cite{bolding2017complementary} considers binary OR activation which can be model by choosing $f(\cdot)$ as a step function, and \cite{pearce2003chemical} employs a linear model for sensor array optimization, i.e., $f(x)=x$. }, namely
\begin{IEEEeqnarray}{lll}  \label{Eq:SoftPlus}
	f(x) = \begin{cases}
		x-x_{\thr} ,\quad &\text{if}\,\,x\geq x_{\thr} \\
		0, &\text{if}\,\,x< x_{\thr},
	\end{cases}
\end{IEEEeqnarray}
where $x_{\thr}$ denotes the receptor activation threshold. The property of inhibition \textbf{P4} can be incorporated in \eqref{Eq:Reception} by assuming that some entries of $\bA$ assume negative values, see \cite[Fig.~6]{pfister2020odorant} for an example characterization of inhibitory/excitatory odor-OR affinity.  A sample construction method for $\bA$ is provided in Section~\ref{sec:A}.  Moreover, $\bn=[n_1,\dots,n_R]^\Trans$ $\in\sRp^R$ denotes the noise vector modeling property \textbf{P5},  where $n_r$ is the aggregated random baseline noise of all type $r$ receptors, which is assumed to be independent of the molecule concentrations $\bx$ and follows a Poisson distribution with mean $\lambda_r$, i.e., $n_r\sim{\rm Pois}(\lambda_r)$ \cite[Remark~21]{jamali2019channel}.  Finally, we note that  the sensitivity and SNR enhancement can be controlled by the relative values of the elements of $\bA$ and $\bn$, i.e., property \textbf{P6}.

\subsection{Sample affinity matrix construction}\label{sec:A} 
A key parameter determining the performance of a CRRA is the affinity matrix $\bA$, which plays a similar role as the measurement matrices  widely used in the CS literature \cite{arjoune2018performance,duarte2011structured}. Instead of focusing on a particular affinity matrix (such as those described in \cite[Fig.~6]{malnic1999combinatorial} and \cite[Fig.~4]{sicard1984receptor}), we follow the  approach that is common in the CS literature and \textit{construct} an affinity matrix that complies with the discussed olfaction properties; please refer to   \cite{arjoune2018performance,duarte2011structured} for an overview of structured/unstructured random/deterministic constructions of measurement matrices. The constructed affinity matrix may be used for guiding the choice of the molecule and receptor types. However, the construction methods discussed in \cite{arjoune2018performance,duarte2011structured} were developed for conventional communication systems and do not meet the specific constraints of  affinity matrices of  olfactory systems, e.g., tuning the relative strengths of activation and inhibition modes of receptors, cf. property \textbf{P1} and \textbf{P4}. Therefore, in the following, we present an example algorithm for the construction of a semi-random measurement matrix that adheres to the specific properties of olfaction discussed in Section~\ref{sec:olfprop}. The proposed design follows the following considerations:
\begin{itemize}
	\item For simplicity, we assume that the maximum entry in each column of $\bA$ is normalized to one, i.e., $\max_r a_{q,r}=1,\,\,\forall q$. In other words, for each type of molecule, the type of receptor that generates the strongest signal has normalized affinity one.   
	\item To account for property \textbf{P1}, we assume that each type of molecule activates $R_{\rm act}$ randomly-chosen receptor types. 
	\item To allow for property \textbf{P4}, we assume that the entries of $\bA$ are randomly drawn from interval $a_{q,r}\in[-a_{\rm inh},1],\,\,\forall q,r$, where  negative values imply molecule inhibition. Here, parameter $a_{\rm inh}\in[0,1]$ controls the strength of the maximum inhibition.
	\item The $q$-th column of $\bA$, denoted by $\ba_{q}\in\sR^{R}$, determines how information regarding the presence of molecule type $q$ is distributed across the receptor array. Therefore, the mutual coherence between any two rows of $\bA$, denoted by $\frac{|\ba_q^\Trans\ba_{q'}|}{\|\ba_q\|\|\ba_{q'}\|}$, should be as small as possible in order to avoid the adoption of molecule types that generate similar activation patterns at the receptor array.
\end{itemize} 
Based on the above considerations, the construction of the affinity matrix $\bA$ is summarized in Algorithm~\ref{Alg:Measurement}, where $\mu_{\rm ch}$ is the maximum coherence allowed between the columns of $\bA$. An example of a constructed affinity matrix with $R=10$ receptor types, $M=20$ molecule types, and parameters $a_{\rm inh}=0.3$, $\mu_{\rm thr}=0.5$, and $R_{\rm act} = 5$ is given in \eqref{Eq:A} on the top of next page.
\begin{figure*}
\begin{IEEEeqnarray}{rll}\label{Eq:A}\fontsize{4.5}{3}\selectfont
	\bA\!=\!
	\arraycolsep=1.3pt\def\arraystretch{2}
	\begin{array}{c}
		\text{R1}\\\text{R2}\\\text{R3}\\\text{R4}\\\text{R5}\\\text{R6}\\\text{R7}\\\text{R8}\\\text{R9}\\\text{R10}
	\end{array}	  	
	\overset{\arraycolsep=6pt\def\arraystretch{2}
		\begin{array}{cccccccccccccccccccc}
			~\text{Q1} & ~\text{Q2} & ~\text{Q3} & ~\text{Q4} & ~\text{Q5} & ~\text{Q6} & ~\text{Q7} & ~\text{Q8} & ~\text{Q9} & \text{Q10} & \text{Q11} & \text{Q12} & \text{Q13} & \text{Q14} & \text{Q15} & \text{Q16} & \text{Q17} & \text{Q18} & \text{Q19} & \text{Q20} 
		\end{array}
	}{ 
		\left[\arraycolsep=1.3pt\def\arraystretch{2}
		\begin{array}{cccccccccccccccccccc}
			0 & 0 & 0 & 0 & 0 & 0.55 & 1 & -0.1 & 0 & 0 & 0 & -0.28 & 0.46 & 0.66 & 1 & 0 & 0.76 & -0.14 & 0 & 0 \\ 
			0 & -0.06 & 0.31 & 0.02 & 1 & 0 & 0 & 0.38 & 0.38 & -0.29 & 0 & 0 & 1 & 0 & 0 & 0 & 0.81 & 0.99 & 0 & 0 \\ 
			0 & 0 & 1 & 0.52 & 0.38 & 0.6 & 0 & -0.11 & 0 & 1 & 0 & 0 & 0 & 0 & 0 & 0.01 & 0.98 & 0 & 0 & 0 \\ 
			1 & 0.41 & 0 & 0 & 0 & 0 & 0.27 & 0 & 0.9 & 0 & -0.25 & 0.65 & 0 & -0.25 & 0 & 0 & 1 & 0 & 1 & 0.76 \\ 
			0 & 0 & 0 & 1 & 0 & 0 & -0.01 & 0 & -0.25 & 0 & 0.71 & -0.17 & 0.73 & 0 & 0.38 & 0 & 0 & -0.1 & 0.88 & 0.79 \\ 
			0.55 & 0.44 & 0.55 & 0 & -0.25 & 0.29 & 0 & 1 & 0 & 0 & 0.31 & 0 & 0 & -0.24 & 0.96 & 0.63 & -0.24 & 0 & 0 & 0 \\ 
			-0.3 & 1 & 0 & 0 & 0.5 & -0.29 & 0 & 0 & 0.33 & 0.6 & 0 & 0 & 0 & 1 & 0.12 & -0.17 & 0 & 0 & -0.07 & 0.75 \\ 
			0 & 0 & 0.62 & 0 & 0 & 0 & 0 & -0.19 & 1 & -0.17 & 1 & 0 & -0.2 & -0.13 & 0.4 & 0.55 & 0 & 0 & 0 & 0.36 \\ 
			-0.08 & 0.67 & 0 & 0 & 0 & 1 & -0.21 & 0 & 0 & 0 & 0.45 & 1 & 0 & 0 & 0 & 0 & 0 & 1 & 0.77 & 0 \\ 
			0.16 & 0 & -0.3 & 0.16 & 0.83 & 0 & 0.89 & 0 & 0 & 0.16 & 0 & 0.84 & -0.18 & 0 & 0 & 1 & 0 & 0 & -0.21 & 1
		\end{array} 
		\right] 
	}	\quad\quad\,\,
\end{IEEEeqnarray}
\hrulefill
\end{figure*}

\begin{algorithm}[t] 
	\caption{\small Example Construction of the Affinity Matrix $\bA$}\scriptsize
	\textbf{input:} $\#$ receptor types $R$, $\#$ molecule types $Q$, inhibition threshold $a_{\rm inh}$, and mutual coherence threshold $\mu_{\rm thr}$ \newline 
	\textbf{output:} Affinity matrix $\bA$
	\begin{algorithmic}[1]\label{Alg:Measurement}
		\FOR{$q=1,\dots,Q$}		
		\STATE Generate $\hat{\ba} = [\hat{a}_1,\dots,\hat{a}_R]^\Trans$ with $R_{\rm act}$ randomly selected entries $\hat{a}_r$ being uniformly distributed RVs in $(0,1]$ and the remaining being zero.
		\STATE Compute $\bar{\ba}= [\bar{a}_1,\dots,\bar{a}_R]^\Trans$ with $\bar{a}_r = \frac{\hat{a}_r}{\max_{r'} \hat{a}_{r'}}(1+a_{\rm inh})-a_{\rm inh}$ if $\hat{a}_r\neq 0$ and $\bar{a}_r =0$ otherwise.
		\STATE Compute $\mu={\max}_{q'<q} \frac{|\bar{\ba}^\Trans\ba_{q'}|}{\|\bar{\ba}\|\|\ba_{q'}\|}$.
		\IF{$q\neq 1$ OR $\mu>\mu_{\rm thr}$}		
		\STATE Go to line~2.
		\ELSE
		\STATE Set $\ba_q=\bar{\ba}$.
		\ENDIF	
		\ENDFOR 		 
	\end{algorithmic}
\end{algorithm}

\section{CS-Based  Recovery Problem}\label{sec:Rx}
In this paper, we assume that the MMSK modulation alphabets employed by the Txs are known to the Rx and focus our attention on the recovery problem.  We refer the interested readers to \cite{jamali2022olfaction}, which is an extended version of this paper, for the design of the MMSK modulation alphabets.

\subsection{Baseline recovery problem} Motivated by the sparsity of $\bx$, we formulate the recovery task as a CS problem. For rigorousness of presentation, we use $\byrv\in\sRp^{R}$ to denote an RV representing the received signal where $\by$ denotes one realization of  $\byrv$. A typical CS problem formulation is to choose the sparsest $\bx$ that reconstructs a signal close to the observation $\by$ based on the adopted measurement model, e.g., \eqref{Eq:Reception}, namely \cite{blumensath2013compressed,defraene2013declipping}
\begin{IEEEeqnarray}{rll}\label{Eq:CS}
	\text{OP0}:	\underset{\bx\in\sR_{\geq0}^{Q}}{\min}\,\, &\|\bx\|_0\nonumber\\
	\text{s.t.}\quad & \big\|\by-\Ex\{\byrv\}\big\|_2^2\leq \epsilon',
\end{IEEEeqnarray}
where $\epsilon'$ is the threshold parameter that controls the reconstruction error and the expectation is with respect to (w.r.t.) noise $\bn$ for given molecule concentration $\bx$. The value of $\epsilon'$ depends on the variance of $\byrv$ for a given $\bx$, which is determined by $\lambda_r$. For simplicity,  throughout this paper, we employ $\epsilon'=\epsilon\bone_R^{\Trans}\Vx\{\byrv\}$, where $\epsilon$ is the normalized reconstruction error threshold.


Problem OP0 faces few shortcomings in general and in particular for the problem under consideration in this paper, which are summarized in the following. \textit{i)} The zero-norm optimization is a non-deterministic polynomial-time (NP) hard problem. This issue is often addressed in the literature by replacing $\|\cdot\|_0$ with its convex one-norm relaxation $\|\cdot\|_1$ \cite{blumensath2013compressed}. \textit{ii)} The constraint in \eqref{Eq:CS} is non-convex due to non-linear receptor activation function $f(\cdot)$. A similar challenge exists for the recovery of clipped audio signals. A convex reformulation of the clipping function was proposed in \cite{defraene2013declipping} via separating the clipped and un-clipped signals. \textit{iii)}  The problem in  \eqref{Eq:CS} treats all possible $\bx$ similarly and does not exploit the fact that depending on the  adopted MMSK modulation alphabets, only certain mixtures may be sent by the Txs.  Next, we reformulate recovery problem OP0 to address these issues.

\subsection{Convex recovery problem}

 To cope with issues \textit{i)} and \textit{ii)}, we adapt the technique used in \cite{defraene2013declipping} to the problem in \eqref{Eq:CS} accounting for the fact that, unlike the Gaussian noise assumed in  \cite{defraene2013declipping}, the interfering noise molecules in our setup are not zero-mean RVs and follow the Poisson statistical model. This leads to:
\begin{IEEEeqnarray}{rl}   \label{Eq:CSprop_x_origninal}
\text{OP1:}\underset{\bx\in\sRp^{Q}}{\min}\,\,\, &\|\bx\|_1 \nonumber\\
\text{s.t.}  \,\,
\text{C1:}\,\, &\big\|\by_{\cA}-\Ex\{\byrv_{\cA}^{\rm lin}\}\big\|_2^2\leq \epsilon\Vx\big\{\bone_{|\cA|}^\Trans\byrv_{\cA}^{\rm lin}\big\} \nonumber\\
\text{C2:}\,\,  & \Ex\{\byrv_{\cAc}^{\rm lin}\} \leq  x_{\thr}\bone_{|\cAc|}+\sqrt{\epsilon\Vx\big\{\byrv_{\cAc}^{\rm lin}\big\}}
\end{IEEEeqnarray}
where  $\cA$ and $\cAc$ are the sets of activated and non-activated receptors, respectively, $\by_{\cA}\in\sRp^{|\cA|}$ collects the measurements from the activated  receptors,  $\byrv^{\rm lin}\triangleq\bA\bx+\bn$ is an RV that is the input to  activation function $f(\cdot)$ in \eqref{Eq:Reception},  $\byrv^{\rm lin}_{\cA}\in\sRp^{|\cA|}$ and $\byrv^{\rm lin}_{\cAc}\in\sRp^{|\cAc|}$ collect the elements of $\byrv^{\rm lin}$ from the activated and non-activated receptors, respectively. Thereby, constraint C1 limits the reconstruction error over the set of activated receptors. Since the noise is random and unknown, the exact condition $\byrv^{\rm lin}_{\cAc}\leq x_{\thr}\bone_{\cAc}$ cannot be applied for the non-activated receptors. Therefore, we consider the statistical constraint in C2 where the normalized parameter $\epsilon$ controls the strictness of this constraint. Evaluating the expectation and  variance in \eqref{Eq:CSprop_x_origninal} using the signal model in \eqref{Eq:Release}-\eqref{Eq:Reception} leads to
\begin{IEEEeqnarray}{rl}   \label{Eq:CSprop_x}
\text{OP1:}\underset{\bx\in\sRp^{Q}}{\min}\,\,\, &\|\bx\|_1 \nonumber\\
\text{s.t.}  \,\,
\text{C1:}\,\, &\|\by_{\cA}-(\bA_{\cA}\bx+(\lambda_r-x_{\thr})\bone_{|\cA|})\|_2^2\leq|\cA|\lambda_r\epsilon \nonumber\\
\text{C2:}\,\,  &\bA_{\cAc}\bx+(\lambda_r-x_{\thr})\bone_{|\cAc|} \leq \sqrt{\lambda_r\epsilon}\bone_{|\cAc|},
\end{IEEEeqnarray}
where  $\bA_{\cA}\in\sR^{|\cA|\times Q}$ and $\bA_{\cAc}\in\sR^{|\cAc|\times Q}$ are matrices containing only the rows of $\bA$ corresponding to activated and non-activated receptors, respectively.

\subsection{Proposed mixture-alphabet aware recovery problem}

Problem OP1 employs the space of molecule concentrations as the signal recovery space.  In order to cope with issue \textit{iii)}, we choose the space of the concentrations of the molecule mixtures as the signal recovery space. In other words, instead of deciding which types of molecules are present around Rx, we directly determine which mixtures are present around Rx. A similar concept is used to describe  mixture identification by natural olfactory systems \cite{thomas2014perception}, which is known as elemental and configural processing of mixtures, where in the former case, a mixture is identified by its constituents, whereas in the latter case, it is identified as a unique quantity.    
To formalize this, let us decompose the \textit{expected} number of received molecule types as $\bar{\bx}=\bV\bM\bs$. Since the release times of the mixtures are random and not known at the Rx, the value of $\bV$ cannot be known. In the following, we assume that the adopted signaling molecule types in each mixture have similar  propagation properties (e.g., diffusion coefficient). Absorbing $\bV$ into $\bs$ for the non-zero entry of $\bs$, we   can write $\bar{\bx}=\bM\bw$, where $\bw\in\sRp^{M}$ is the vector of \textit{expected} numbers of molecule types that reach the Rx for the released mixture.
Recall that for a given $\bar{\bx}$, the actual number of received molecules is an RV, cf. \eqref{Eq:Propagation}. Therefore, for rigorousness of presentation, we use $\bxrv\in\sRp^{Q}$ to denote an RV representing the vector of received numbers of different molecule types where  $\bx$ denotes one realization of  $\bxrv$.  Using these notations, we propose the following recovery~problem: 
\begin{IEEEeqnarray}{rll}  \label{Eq:CSprop_mix}
		\text{OP2}:		\underset{\bx\in\sRp^{Q},\bw\in\sRp^{M}}{\min}\quad &\|\bw\|_1 \nonumber\\
		\text{s.t.}  \,\,
		\text{C1}, \text{C2}, 
		\text{C3:}\,\,& |\bx-\Ex\{\bxrv\}|^2\leq \delta\Vx\{\bxrv\},
\end{IEEEeqnarray}
where constraint C3 controls the maximum deviation of the estimated molecule concentration $\bx$ from  the mean of $\bxrv$. The amount of allowable deviation, in general, depends on the variance of $\bxrv$, which, for the Poisson model in \eqref{Eq:Propagation}, is equal to its mean, i.e., $\Ex\{\bxrv\}=\Vx\{\bxrv\}=\bar{\bx}=\bM\bw$. Therefore, we parameterized the maximum deviation as $\delta\Vx\{\bxrv\}$, where $\delta$ is a constant threshold to control the deviation. 
The optimal values of $\delta$ and $\epsilon$ are numerically determined for different setups in Section~\ref{sec:sim}. 
The optimization problem in \eqref{Eq:CSprop_mix} is convex and can be solved by standard optimization toolboxes, e.g., CVX. 

We assume that each Tx releases only one mixture at a time and that, due to the random access by the Txs, the probability that two Txs release mixtures simultaneously is negligible. Therefore, we employ a simple peak detector to determine which molecule mixture is present at each sampling~time:
\begin{IEEEeqnarray}{lll}  \label{Eq:ThrDec}
	\hat{s}_{m} = \begin{cases}
		1,	& \mathrm{if}\,\, m=\underset{m'\in\sM}{\argmax}\,\hat{w}_{m'}\\
		0, & \mathrm{otherwise},
	\end{cases}
\end{IEEEeqnarray}
where  $\hat{\bw}=[\hat{w}_1,\dots,\hat{w}_M]^\Trans$ denotes the solution of OP2.
\vspace{-0.05cm}

\section{Simulation Results}\label{sec:sim}

Unless stated otherwise, the values of the system parameters are chosen as follow: $K=4$, $R=10$, $Q=20$, $N_{\rls}=5\times 10^3$, $v_q=0.01$, $\lambda_r=10$, $x_{\rm thr}=5$, and $\bA$ is chosen from  Eq. \eqref{Eq:A}. We assume that each Tx spontaneously accesses the channel and sends one mixture corresponding to one message. The transmission, propagation, and reception of molecules follow  \eqref{Eq:Release}, \eqref{Eq:Propagation}, and \eqref{Eq:Reception}, respectively. We adopt the receptor-affinity matrix in \eqref{Eq:A} and the MMSK alphabets are chosen from \cite[Table~II]{jamali2022olfaction}, where for all adopted mixtures, the molecule distribution is assumed to be uniform. Finally, we adopt OP2 and \eqref{Eq:ThrDec} for data recovery.

\begin{figure}[t]
	\centering
\includegraphics[width=0.8\columnwidth, angle =0]{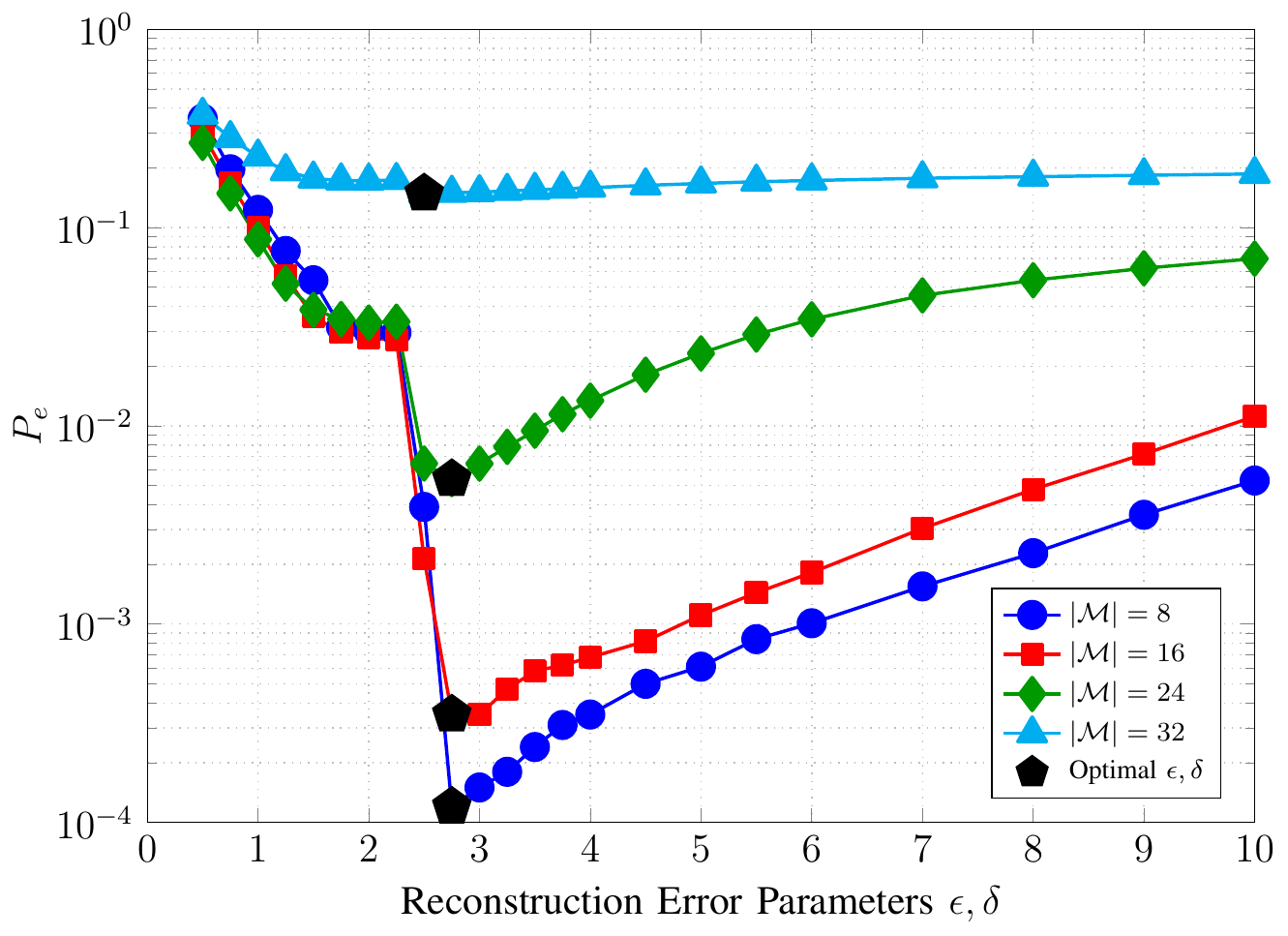}
\vspace{-0.2cm}
\caption{Error probability $P_e$ vs. reconstruction error parameters $\epsilon,\delta$ for different total numbers of mixtures $|\cM|$ used by all Txs.\vspace{-0.3cm}} \label{Fig:Pe_known} 
\end{figure}

In Fig.~\ref{Fig:Pe_known}, we plot the error probability vs. the reconstruction error parameter for different numbers of mixtures used by all Txs $|\sM|\triangleq K M_{\rm tx}$, where $M_{\rm tx}$  is the MMSK alphabet size for each Tx. Since both reconstruction error parameters (i.e., $\epsilon$ and $\delta$) are similarly normalized, cf. \eqref{Eq:CSprop_mix}, we assume identical values for them, i.e., $\epsilon=\delta$. We observe from Fig.~\ref{Fig:Pe_known} that for each curve, there exists an optimal value for the reconstruction error parameter that minimizes the error probability. This is due to the fact that while large  $\epsilon,\delta$ lead to inaccurate estimates of $\bx$ and $\bw$ (i.e., underfitting), small  $\epsilon,\delta$ lead to an infeasible recovery problem and overfitting. Moreover, it can be observed from this figure that as the number of mixtures increases, the error probability increases. This behavior stems from the fact that by increasing $|\sM|$, the minimum distance among  mixtures in the receptor array space reduces which constitutes the bottleneck for  the recovery performance. { We note that for conventional MSK modulation, where each message is represented by one type of molecule, the complexity of the DREAM Rx linearly scales with the number of messages. In contrast, for the proposed MMSK modulation and the CRRA Rx, the number of messages, $|\sM|$, may vary for a fixed Rx and be even much larger than the number of receptor types, i.e., $|\sM|\gg R$,  at the expense of a reduction in  detection reliability, of course. In other words, the proposed MC system offers a tunable tradeoff between the system throughput, $|\sM|$, and the detection reliability, $P_e$, for a fixed Rx architecture.}

\begin{figure}[t]
	\centering
	\includegraphics[width=0.8\columnwidth, angle =0]{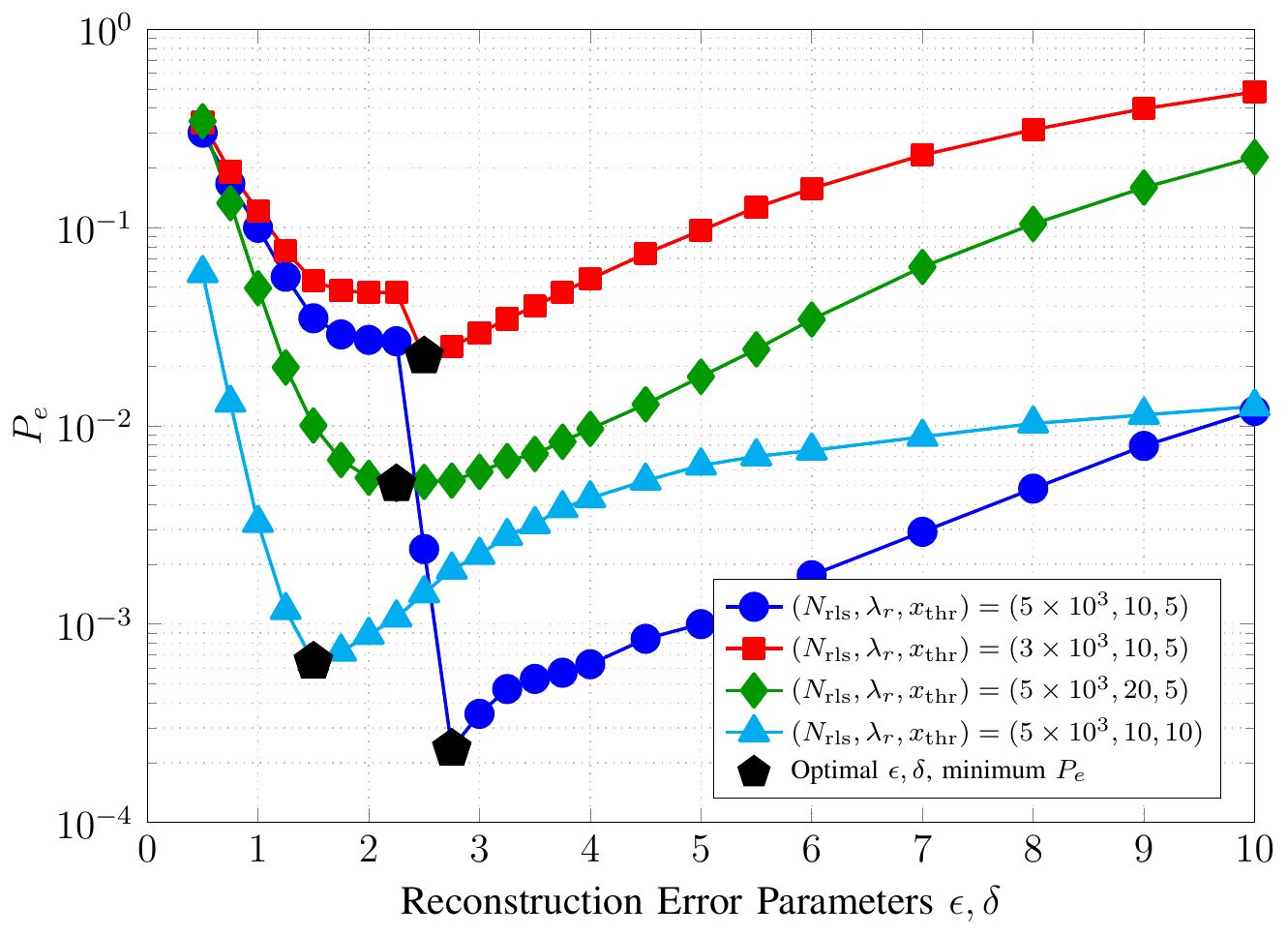}
	\vspace{-0.2cm}
	\caption{Error probability $P_e$ vs. reconstruction error parameters $\epsilon,\delta$ for $|\sM|=16$ and different sets of  parameters $(N_{\rm rls},\lambda_r,x_{\rm thr})$. \vspace{-0.3cm}} \label{Fig:Pe_param}  
\end{figure}

Next, we investigate the impact of system parameters on the recovery performance. In particular, in Fig.~\ref{Fig:Pe_param}, we show the error probability vs. the reconstruction error parameters for $|\sM|=16$ and different values of system parameters $(N_{\rm rls},\lambda_r,x_{\rm thr})$. It can be observed from this figure that an increase in the number of  released molecules, $N_{\rm rls}$, significantly improves the recovery performance. There are two reasons for this performance improvement. First, a higher concentration implies a higher SNR and reduces the impact of noise. Second, higher concentrations of molecules  activate even receptors with low affinity, which implies that more knowledge about the present mixtures is contained in the receptor array signal.  This argument is further confirmed by Fig.~\ref{Fig:Pe_param} as the recovery performance improves for smaller  noise means $\lambda_r$ and smaller activation thresholds~$x_{\rm thr}$.  

\section{Conclusions}
We have proposed and studied  a novel olfaction-inspired  synthetic MC system. In the proposed MC system, Txs employ mixtures of molecules as information carriers (i.e., MMSK modulation) and the Rx is equipped with an array of cross-reactive receptors in order to identify the transmitted mixtures. We first introduced an end-to-end MC channel model that accounts for the key properties of olfaction.  Subsequently, we formulated the molecule mixture recovery as a convex CS problem, i.e., OP2, which explicitly takes into account knowledge of the MMSK modulation alphabets employed by the Txs. Our simulation results revealed the efficiency of the proposed CS problem for the recovery of the  MMSK modulated signals and quantified the system performance for various system parameters. 

\section*{Acknowledgment}
Vahid Jamali’s work was funded by the Deutsche Forschungsgemeinschaft (DFG -- German Research Foundation) under project number JA 3104/1-1.

\bibliographystyle{IEEEtran}
\bibliography{References}

\begin{thebibliography}{10}
\providecommand{\url}[1]{#1}
\csname url@samestyle\endcsname
\providecommand{\newblock}{\relax}
\providecommand{\bibinfo}[2]{#2}
\providecommand{\BIBentrySTDinterwordspacing}{\spaceskip=0pt\relax}
\providecommand{\BIBentryALTinterwordstretchfactor}{4}
\providecommand{\BIBentryALTinterwordspacing}{\spaceskip=\fontdimen2\font plus
\BIBentryALTinterwordstretchfactor\fontdimen3\font minus
  \fontdimen4\font\relax}
\providecommand{\BIBforeignlanguage}[2]{{%
\expandafter\ifx\csname l@#1\endcsname\relax
\typeout{** WARNING: IEEEtran.bst: No hyphenation pattern has been}%
\typeout{** loaded for the language `#1'. Using the pattern for}%
\typeout{** the default language instead.}%
\else
\language=\csname l@#1\endcsname
\fi
#2}}
\providecommand{\BIBdecl}{\relax}
\BIBdecl

\bibitem{jamali2019channel}
V.~Jamali, A.~Ahmadzadeh, W.~Wicke, A.~Noel, and R.~Schober, ``Channel modeling
  for diffusive molecular communication--{A} tutorial review,'' \emph{Proc. of
  the IEEE}, vol. 107, no.~7, pp. 1256--1301, 2019.

\bibitem{kuscu2019transmitter}
M.~Kuscu, E.~Dinc, B.~A. Bilgin, H.~Ramezani, and O.~B. Akan, ``Transmitter and
  receiver architectures for molecular communications: {A} survey on physical
  design with modulation, coding, and detection techniques,'' \emph{Proc. of
  the IEEE}, vol. 107, no.~7, pp. 1302--1341, 2019.

\bibitem{kuran2011modulation}
M.~S. Kuran, H.~B. Yilmaz, T.~Tugcu, and I.~F. Akyildiz, ``Modulation
  techniques for communication via diffusion in nanonetworks,'' in \emph{Proc.
  IEEE Int. Conf. Commun.}, 2011, pp. 1--5.

\bibitem{jamali2018diffusive}
V.~Jamali, N.~Farsad, R.~Schober, and A.~Goldsmith, ``Diffusive molecular
  communications with reactive molecules: {Channel} modeling and signal
  design,'' \emph{IEEE Trans. Molecular, Biol., and Multi-Scale Commun.},
  vol.~4, no.~3, pp. 171--188, 2018.

\bibitem{chen2020generalized}
X.~Chen, Y.~Huang, L.-L. Yang, and M.~Wen, ``Generalized molecular-shift keying
  ({GMoSK}): {Principles} and performance analysis,'' \emph{IEEE Trans.
  Molecular, Biol., and Multi-Scale Commun.}, vol.~6, no.~3, pp. 168--183,
  2020.

\bibitem{tisch2010nanomaterials}
U.~Tisch and H.~Haick, ``Nanomaterials for cross-reactive sensor arrays,''
  \emph{MRS Bulletin}, vol.~35, no.~10, pp. 797--803, 2010.

\bibitem{buck2005unraveling}
L.~B. Buck, ``Unraveling the sense of smell ({Nobel} lecture),''
  \emph{Angewandte Chemie International Edition}, vol.~44, no.~38, pp.
  6128--6140, 2005.

\bibitem{su2009olfactory}
C.-Y. Su, K.~Menuz, and J.~R. Carlson, ``Olfactory perception: {Receptors},
  cells, and circuits,'' \emph{Cell}, vol. 139, no.~1, pp. 45--59, 2009.

\bibitem{kaupp2010olfactory}
U.~B. Kaupp, ``Olfactory signalling in vertebrates and insects: {Differences}
  and commonalities,'' \emph{Nature Reviews Neuroscience}, vol.~11, no.~3, pp.
  188--200, 2010.

\bibitem{thomas2014perception}
T.~Thomas-Danguin, C.~Sinding, S.~Romagny, F.~El~Mountassir, B.~Atanasova,
  E.~Le~Berre, A.-M. Le~Bon, and G.~Coureaud, ``The perception of odor objects
  in everyday life: {A} review on the processing of odor mixtures,''
  \emph{Frontiers in Psychology}, vol.~5, p. 504, 2014.

\bibitem{jamali2022olfaction}
\BIBentryALTinterwordspacing
V.~Jamali, H.~M. Loos, A.~Buettner, R.~Schober, and H.~V. Poor,
  ``Olfaction-inspired {MCs}: {Molecule} mixture shift keying and
  cross-reactive receptor arrays,'' \emph{Submitted to IEEE Trans. Commun.},
  2022. [Online]. Available: \url{https://arxiv.org/pdf/2203.04225.pdf}
\BIBentrySTDinterwordspacing

\bibitem{meng2012mimo}
L.-S. Meng, P.-C. Yeh, K.-C. Chen, and I.~F. Akyildiz, ``{MIMO} communications
  based on molecular diffusion,'' in \emph{Proc. IEEE Global Commun. Conf.},
  2012, pp. 5380--5385.

\bibitem{koo2016molecular}
B.-H. Koo, C.~Lee, H.~B. Yilmaz, N.~Farsad, A.~Eckford, and C.-B. Chae,
  ``Molecular {MIMO}: {From} theory to prototype,'' \emph{IEEE J. Select. Areas
  in Commun.}, vol.~34, no.~3, pp. 600--614, 2016.

\bibitem{mori1995molecular}
K.~Mori and Y.~Yoshihara, ``Molecular recognition and olfactory processing in
  the mammalian olfactory system,'' \emph{Progress in Neurobiology}, vol.~45,
  no.~6, pp. 585--619, 1995.

\bibitem{bolding2017complementary}
K.~A. Bolding and K.~M. Franks, ``Complementary codes for odor identity and
  intensity in olfactory cortex,'' \emph{Elife}, vol.~6, p. e22630, 2017.

\bibitem{hallem2004molecular}
E.~A. Hallem, M.~G. Ho, and J.~R. Carlson, ``The molecular basis of odor coding
  in the {Drosophila} antenna,'' \emph{Cell}, vol. 117, no.~7, pp. 965--979,
  2004.

\bibitem{pfister2020odorant}
P.~Pfister, B.~C. Smith, B.~J. Evans, J.~H. Brann, C.~Trimmer, M.~Sheikh,
  R.~Arroyave, G.~Reddy, H.-Y. Jeong, D.~A. Raps \emph{et~al.}, ``Odorant
  receptor inhibition is fundamental to odor encoding,'' \emph{Current
  Biology}, vol.~30, no.~13, pp. 2574--2587, 2020.

\bibitem{wilson2013early}
R.~I. Wilson, ``Early olfactory processing in {Drosophila: Mechanisms} and
  principles,'' \emph{Annual Review of Neuroscience}, vol.~36, pp. 217--241,
  2013.

\bibitem{laurent1999systems}
G.~Laurent, ``A systems perspective on early olfactory coding,''
  \emph{Science}, vol. 286, no. 5440, pp. 723--728, 1999.

\bibitem{malnic1999combinatorial}
B.~Malnic, J.~Hirono, T.~Sato, and L.~B. Buck, ``Combinatorial receptor codes
  for odors,'' \emph{Cell}, vol.~96, no.~5, pp. 713--723, 1999.

\bibitem{sicard1984receptor}
G.~Sicard \emph{et~al.}, ``Receptor cell responses to odorants: {Similarities}
  and differences among odorants,'' \emph{Brain Research}, vol. 292, no.~2, pp.
  283--296, 1984.

\bibitem{qin2019optimal}
S.~Qin, Q.~Li, C.~Tang, and Y.~Tu, ``The optimal odor-receptor interaction
  network is sparse in olfactory systems: {Compressed} sensing by nonlinear
  neurons with a finite dynamic range,'' \emph{bioRxiv}, p. 464875, 2019.

\bibitem{pearce2003chemical}
T.~C. Pearce and M.~Sanchez-Montanes, ``Chemical sensor array optimization:
  {Geometric} and information theoretic approaches,'' \emph{Handbook of
  Artificial Olfaction Machines}, pp. 347--376, 2003.

\bibitem{arjoune2018performance}
Y.~Arjoune, N.~Kaabouch, H.~El~Ghazi, and A.~Tamtaoui, ``A performance
  comparison of measurement matrices in compressive sensing,'' \emph{Int. J.
  Commun. Syst.}, vol.~31, no.~10, p. e3576, 2018.

\bibitem{duarte2011structured}
M.~F. Duarte and Y.~C. Eldar, ``Structured compressed sensing: {From} theory to
  applications,'' \emph{IEEE Trans. Sig. Process.}, vol.~59, no.~9, pp.
  4053--4085, 2011.

\bibitem{blumensath2013compressed}
T.~Blumensath, ``Compressed sensing with nonlinear observations and related
  nonlinear optimization problems,'' \emph{IEEE Trans. Inf. Theory}, vol.~59,
  no.~6, pp. 3466--3474, 2013.

\bibitem{defraene2013declipping}
B.~Defraene, N.~Mansour, S.~De~Hertogh, T.~Van~Waterschoot, M.~Diehl, and
  M.~Moonen, ``Declipping of audio signals using perceptual compressed
  sensing,'' \emph{IEEE Trans. Audio, Speech, Language Process.}, vol.~21,
  no.~12, pp. 2627--2637, 2013.

\end{thebibliography}

\end{document}